\documentclass[final,1p,times]{elsarticle} 

\usepackage{graphicx} 
\usepackage{amssymb} 
\usepackage{amsthm} 
\usepackage{lineno} 

\journal{Nuclear Physics A} 
\begin{document} 

\begin{frontmatter} 


\title{From 0 to 5000 in $2\times 10^{-24}$ seconds:\\
Entropy production in relativistic heavy-ion collisions}

\author{R.J.~Fries$^{\rm a,b}$, T.~Kunihiro$^{\rm c}$, B.~M\"uller$^{\rm d}$, A.~Ohnishi$^{\rm e}$ and A.~Sch\"afer$^{\rm f}$}

\address{$^{\rm a}$Department of Physics, Texas A{\&}M University, College Station, TX 77843, USA\\
$^{\rm b}$RIKEN/BNL Research Center, Brookhaven National Laboratory, Upton, NY 11973, USA\\
$^{\rm c}$Department of Physics, Kyoto University, Sakyo-ku, Kyoto 606-8502, Japan\\
$^{\rm d}$Department of Physics, Duke University, Durham, NC 27708, USA\\
$^{\rm e}$Yukawa Institute of Theoretical Physics, Kyoto University, Kyoto 606-8502, Japan\\
$^{\rm f}$Institut f\"ur Theoretische Physik, Universit\"at Regensburg, D-93040 Regensburg, Germany}

\begin{abstract} 
We review what is known about the contributions to the final entropy from the different stages of a relativistic nuclear collision, including recent results on the decoherence entropy and the entropy produced during the hydrodynamic phase by viscous effects. We then present a general framework, based on the Husimi distribution function, for the calculation of entropy growth in quantum field theories, which is applicable to the earliest (``glasma'') phase of the collision during which most of the entropy is generated. The entropy calculated from the Husimi distribution exhibits linear growth when the quantum field contains unstable modes and is asymptotically equal to the Kolmogorov-Sina\"i (KS) entropy. We outline how the approach can be used to investigate the problem of entropy production in a relativistic heavy-ion reaction from first principles.
\end{abstract} 

\end{frontmatter} 


\section{Overview}\label{sec:intro}

\begin{figure}[ht]
\centering
\includegraphics[width=0.8\textwidth]{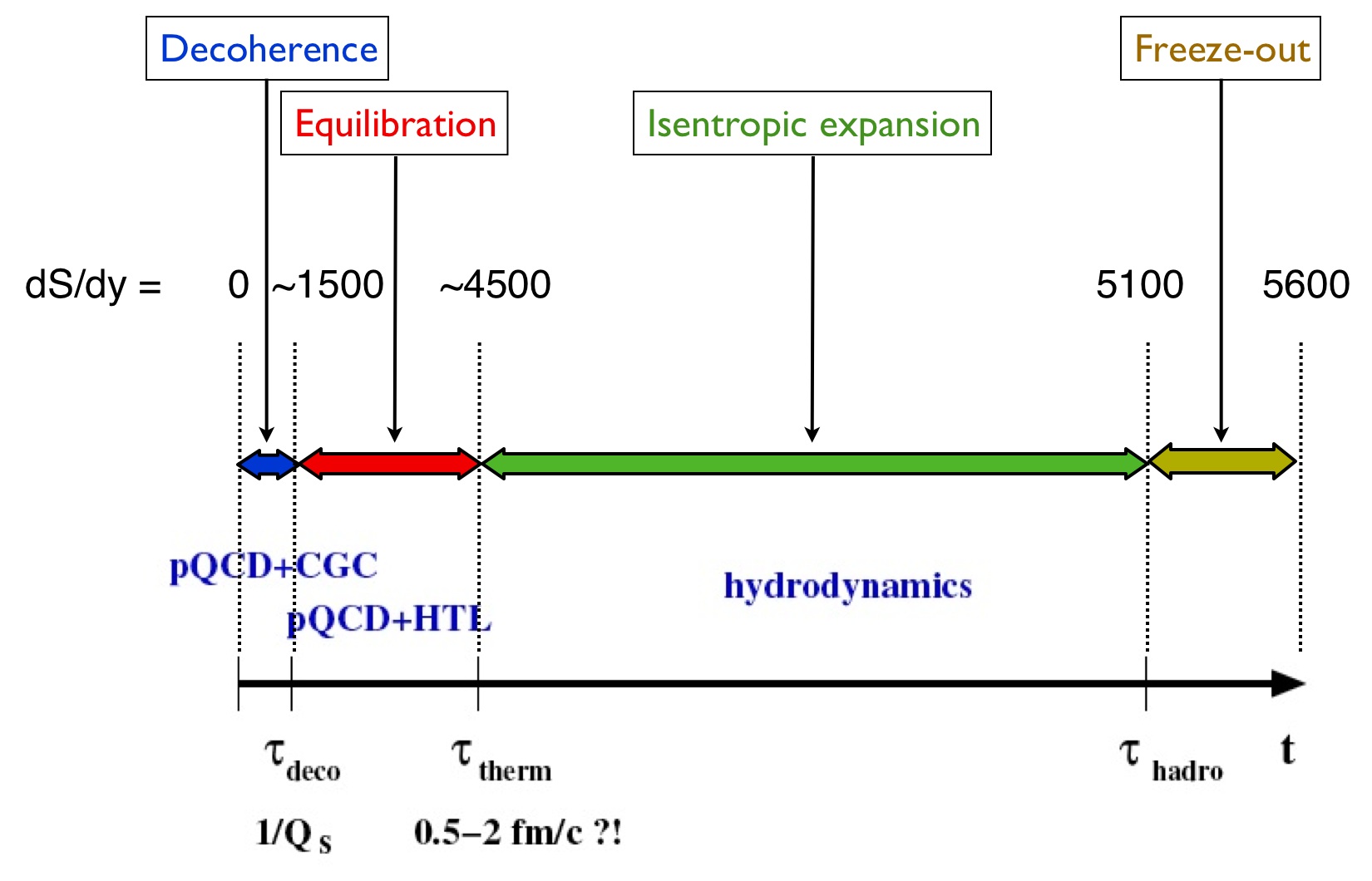}
\caption[]{Entropic history of a central Au+Au collision at top RHIC energy. The values of $dS/dy$ indicate the entropy per unit rapidity reached at the end of various collision stages based on experimental data and model estimates.}
\label{fig:BM1}
\end{figure}

The agreement of hydrodynamic calculations of the flow anisotropy of the matter produced in nuclear collisions at RHIC with the elliptic flow measurements rests on the assumption of an early equilibration of the matter on a time-scale of the order of 1 fm/c. An important problem in the description of relativistic heavy ion reactions is thus to understand how the produced matter equilibrates so quickly. From a thermodynamic standpoint, this question can be answered by studying when and how entropy is created in the reaction. One can distinguish five different stages of entropy production:
\begin{enumerate}
\setlength{\itemsep}{0pt}
\item Decoherence of the initial nuclear wave functions;
\item Thermalization of the partonic plasma (``glasma'');
\item Dissipation due to shear viscosity in the hydrodynamic expansion;
\item Hadronization accompanied by large bulk viscosity;
\item Viscous hadronic freeze-out.
\end{enumerate}
What is known about the contribution of the different stages to the final entropy is depicted in Fig. \ref{fig:BM1}.
First, we note that the final entropy per unit rapidity $dS/dy$ is one of the best known quantities in relativistic heavy ion physics. In the case of Au+Au collisions at RHIC $dS/dy$ at freeze-out can be determined from an analysis of the final hadron spectra in combination with the information on the source radius derived from identical particle (HBT) correlations \cite{Pal:2003rz}. The slightly extrapolated result for the 6\% most central Au+Au collisions at $\sqrt{s_{\rm NN}}= 200$ GeV is $(dS/dy)_{\rm f}=5600 \pm 500$ at midrapidity \cite{Muller:2005en}. Alternatively, the final entropy can be deduced from the measured hadron abundances, combined with the calculated entropy per particle for a hadron gas in chemical equilibrium, which yields the result $(dS/dy)_{\rm ch}=5100 \pm 200$ \cite{Muller:2005en}. The 10 percent difference can be attributed to the entropy production during the hadronic freeze-out and reflects the significant viscosity of a hadronic gas \cite{Demir:2008tr}.

\section{Mechanisms of entropy production}\label{sec:mech}

Broadly speaking, we can distinguish between entropy production by decoherence of the initial quantum state, by the equilibration process, and by dissipative processes in the hydrodynamic expansion phase. Starting with decoherence, the loss of coherence is measured by the decay in off-diagonal elements of the density matrix $\rho$. A practical way of investigating this is to calculate the decay rate of the quantity 
${\rm Tr}\rho^2/[{\rm Tr}\rho]^2$. 
The time scale of the decoherence of the initial nuclear wavefunction has been studied by this method in the color glass condensate model with the result that the characteristic decoherence time is \cite{Muller:2005yu,Fries:2008vp}
$\tau_{\rm deco} = c Q_s^{-1}$,
where $Q_s$ is the nuclear saturation scale, and $c$ denotes a calculable constant close to unity. While this result is expected on dimensional grounds, the fact that $c<1$ is important because it shows that  decoherence occurs over a time of less than 0.2 fm/c. The amount of entropy created from loss of coherence alone can be estimated by multiplying the number of causally disconnected transverse domains in a central Au+Au collision by the entropy $S_{\rm deco} \approx (\ln 2\pi\bar{n} + 1)/2$ obtained from the complete decoherence of a coherent quantum state with average occupancy $\bar{n}$. Accounting for a longitudinal coherence length $\Delta y \approx 1/\alpha_s$ and using $\bar{n}\approx 1/\alpha_s$ one obtains \cite{Fries:2008vp,Muller:2003cr}:
\begin{equation}
(dS/dy)_{\rm deco} \approx \frac{1}{2} Q_s^2 R^2 \alpha_s (\ln 2\pi/\alpha_s +1) \approx 1,500
\label{eq:dSdy}
\end{equation}
at midrapidity for a central Au+Au collision, roughly one quarter of the final entropy. 

The other stage which is reasonably well understood is the hydrodynamic expansion stage. RHIC data on the elliptic flow of hadrons in non-central collisions indicate that the shear viscosity during this phase is small. The bounds on the shear viscosity $\eta$ compatible with the RHIC data lie characteristically in the range $\eta/s \leq (2-3)/4\pi$ \cite{Romatschke:2007mq,Gavin:2007zza,Luzum:2008cw,Song:2008hj,Lacey:2009xx}. Such a small viscosity cannot contribute much to the entropy of the expanding fluid. If one sets the start of the hydrodynamic era at $\tau_{\rm th}\approx 1$ fm/c, where $\tau_{\rm th}$ denotes the thermalization time, one-dimensional solutions of boost invariant hydrodynamics show that the the entropy of the fluid increases by about 15 percent between $\tau_{\rm th}$ and the time of hadronization \cite{Fries:2008ts}. This implies that the entropy at thermalization is approximately $(dS/dy)_{\rm th} \approx 4,500$. 

\section{Entropy growth rate}\label{sec:Husimi}

These considerations tell us that at least half of the final entropy must be generated during the thermalization process. If the quark-gluon plasma were a weakly coupled system, the growth of the entropy during this stage could be calculated from a partonic Boltzmann equation. However, the small shear entropy and other observations from the RHIC experiments, such as jet quenching, indicate that the plasma is strongly coupled. In addition, it is thought that the initial state (the ``glasma'' \cite{Lappi:2006fp,Romatschke:2006nk}), as well as the pre-equilibrium quark-gluon plasma \cite{Mrowczynski:2005ki}, are characterized by strong, partially coherent color fields. Any formalism that hopes to describe entropy production during this early stage must, therefore, be able to describe the growing complexity of a quantum system making the transition from a regime  of field dominance to the hydrodynamic regime. We have recently proposed such a general formalism \cite{Kunihiro:2008gv}.

The idea is basically simple: the growing entropy measures the increasing intrinsic complexity of the quantum state of the system after appropriate coarse graining. The problem is how to impose a minimal amount of course graining without assuming the answer. The solution to this problem dates back to Husimi \cite{Husimi:1940}. The {\em Husimi distribution} is a convolution of the Wigner function $W(p,x,t)$ of the system with a minimum-uncertainty Gaussian wave packet:
\begin{equation}
H_\Delta(p,x;t) = \int \frac{dp'\, dx'}{\pi\hbar}
\exp\left( - \frac{1}{\hbar\Delta}(p-p')^2 - \frac{\Delta}{\hbar}(x-x')^2 \right) W(p',x';t).
\label{eq:Husimi}
\end{equation}
Here $x$ and $p$ stand for all ``position'' and `` momentum'' variables characterizing the system, which may include particle positions as well as field amplitudes (see \cite{Mrowczynski:1994nf} for the definition of the Wigner functional for quantum fields). The Husimi distribution depends on the squeezing parameter $\Delta$, which measures the ratio of position and momentum uncertainty. Its value can be chosen at liberty, but once fixed, will not change with time. The Husimi distribution can be understood as a coarse grained phase space distribution of the quantum system, where the coarse graining corresponds to the projection on a coherent state of the quantum system. We recall that coherent states are the closest quantum analogues of classical systems compatible with the uncertainty relation. Because the Husimi distribution can be shown to be positive (semi-)definite, it permits the definition of a coarse-grained entropy, first introduced by Wehrl 
\cite{Wehrl:1978zz}:
\begin{equation}
S_{\rm H,\Delta}(t) = - \int \frac{dp\, dx}{2\pi\hbar} H_\Delta(p,x;t) \ln H_\Delta(p,x;t).
\label{eq:SH}
\end{equation}
Quantum systems containing unstable modes, i.e.\ modes with an exponentially growing amplitude, have a linearly growing Husimi-Wehrl entropy \cite{Kunihiro:2008gv}. The entropy growth rate is independent of the squeezing parameter $\Delta$ and given by the sum of the exponential growth rates of all unstable modes. In classical dynamical systems, this quantity is known as the Kolmogorov-Sina\"i entropy, or KS entropy, and defined as the sum over all positive Lyapunov exponents $\lambda_k$ of the system:
$dS_{\rm H,\Delta}/dt \longrightarrow S_{\rm KS} 
= \sum_{k} \lambda_k \, \theta(\lambda_k).$
The KS entropy is understood to be a measure of the growth rate of the coarse grained entropy of a dynamical system starting from a configuration far away from equilibrium, after an initial start-up phase during which unstable fluctuations grow to dominance and before it gets too close to its micro-canonical equilibrium \cite{Latora:1999}. 

\section{Entropy growth in QCD}

The Husimi-Wehrl entropy provides the basis for the formulation of a comprehensive approach to entropy production in heavy-ion collisions. The lattice regularized Yang-Mills equations describing the dynamics of classical color fields are known to be strongly chaotic \cite{Muller:1992iw,Biro:1993qc}, and the KS-entropy of the classical Yang-Mills field was shown to be a thermodynamically extensive quantity \cite{Bolte:1999th}. The lattice approach can be extended to include Gaussian fluctuations \cite{Gong:1993fz}. A calculation of entropy production using this approach with a random initial conditions is presently underway \cite{project}, and it would be interesting to evaluate the evolution with the initial quantum fluctuations around classical glue fields in the colliding nuclei \cite{Fukushima:2006ax}.

\section*{Acknowledgments} 

This work was supported in part by grants from the U.S. Department of Energy and the German BMBF, and by the Yukawa International Program for Quark-Hadron Sciences at the Yukawa Institute for Theoretical Physics, Kyoto University, Japan. RJF acknowledges partial support by RIKEN/BNL Research Center.


\begin{thebibliography}{00}  
   
%
\bibitem{Pal:2003rz}
  S.~Pal and S.~Pratt,
  Phys.\ Lett.\  B {\bf 578}, 310 (2004)
  [arXiv:nucl-th/0308077].
%
\bibitem{Muller:2005en}
  B.~M\"uller and K.~Rajagopal,
  Eur.\ Phys.\ J.\  C {\bf 43}, 15 (2005)
  [arXiv:hep-ph/0502174].
%
\bibitem{Demir:2008tr}
  N.~Demir and S.~A.~Bass,
  Phys.\ Rev.\ Lett.\  {\bf 102}, 172302 (2009)
  [arXiv:0812.2422 [nucl-th]].
%
\bibitem{Muller:2005yu}
  B.~M\"uller and A.~Sch\"afer,
  Phys.\ Rev.\  C {\bf 73}, 054905 (2006)
  [arXiv:hep-ph/0512100].
%
\bibitem{Fries:2008vp}
  R.~J.~Fries, B.~M\"uller and A.~Sch\"afer,
  Phys.\ Rev.\  C {\bf 79}, 034904 (2009)
  [arXiv:0807.1093 [nucl-th]].
%
\bibitem{Muller:2003cr}
  B.~M\"uller and A.~Sch\"afer,
  arXiv:hep-ph/0306309.
%
\bibitem{Romatschke:2007mq}
  P.~Romatschke and U.~Romatschke,
  Phys.\ Rev.\ Lett.\  {\bf 99}, 172301 (2007)
  [arXiv:0706.1522 [nucl-th]].
%
\bibitem{Gavin:2007zza}
  S.~Gavin and M.~Abdel-Aziz,
  J.\ Phys.\ G {\bf 34}, S835 (2007).
%
\bibitem{Luzum:2008cw}
  M.~Luzum and P.~Romatschke,
  Phys.\ Rev.\  C {\bf 78}, 034915 (2008)
  [arXiv:0804.4015 [nucl-th]].
%
\bibitem{Song:2008hj}
  H.~Song and U.~W.~Heinz,
  arXiv:0812.4274 [nucl-th].
%
\bibitem{Lacey:2009xx}
  R.~A.~Lacey, A.~Taranenko and R.~Wei,
  arXiv:0905.4368 [nucl-ex].
%
\bibitem{Fries:2008ts}
  R.~J.~Fries, B.~M\"uller and A.~Sch\"afer,
  Phys.\ Rev.\  C {\bf 78}, 034913 (2008)
  [arXiv:0807.4333 [nucl-th]].
%
\bibitem{Lappi:2006fp}
  T.~Lappi and L.~McLerran,
  Nucl.\ Phys.\  A {\bf 772}, 200 (2006)
  [arXiv:hep-ph/0602189].
%
\bibitem{Romatschke:2006nk}
  P.~Romatschke and R.~Venugopalan,
  Phys.\ Rev.\  D {\bf 74}, 045011 (2006)
  [arXiv:hep-ph/0605045].
%
\bibitem{Mrowczynski:2005ki}
  St.~Mr\'owczy\'nski,
  Acta Phys.\ Polon.\  B {\bf 37}, 427 (2006)
  [arXiv:hep-ph/0511052].
%
\bibitem{Kunihiro:2008gv}
  T.~Kunihiro, B.~M\"uller, A.~Ohnishi and A.~Sch\"afer,
  Prog.\ Theor.\ Phys.\  {\bf 121}, 555 (2009)
  [arXiv:0809.4831 [hep-ph]].
%
\bibitem{Husimi:1940}
  K.~Husimi,
  Proc.\ Phys.\ Math.\ Soc.\ Jpn.\ {\bf 22}, 246 (1940).
%
\bibitem{Mrowczynski:1994nf}
  St.~Mr\'owczy\'nski and B.~M\"uller,
  Phys.\ Rev.\  D {\bf 50}, 7542 (1994)
  [arXiv:hep-th/9405036].
%
\bibitem{Wehrl:1978zz}
  A.~Wehrl,
  Rev.\ Mod.\ Phys.\  {\bf 50}, 221 (1978).
%
\bibitem{Latora:1999}
  V.~Latora and M.~Baranger,
  Phys.\ Rev.\ Lett.\ {\bf 82}, 520 (1999).
%
\bibitem{Muller:1992iw}
  B.~M\"uller and A.~Trayanov,
  Phys.\ Rev.\ Lett.\  {\bf 68}, 3387 (1992).
%
\bibitem{Biro:1993qc}
  T.~S.~Bir\'o, C.~Gong, B.~M\"uller and A.~Trayanov,
  Int.\ J.\ Mod.\ Phys.\  C {\bf 5}, 113 (1994).
%
\bibitem{Bolte:1999th}
  J.~Bolte, B.~M\"uller and A.~Sch\"afer,
  Phys.\ Rev.\  D {\bf 61}, 054506 (2000).
%
\bibitem{Gong:1993fz}
  C.~Gong, B.~M\"uller and T.~S.~Bir\'o,
  Nucl.\ Phys.\  A {\bf 568}, 727 (1994).
%
\bibitem{project}
  T.~Kunihiro, B.~M\"uller, A.~Ohnishi, A.~Sch\"afer, T.~Takahashi, and A.~Yamamoto,
  work in progress.
%
\bibitem{Fukushima:2006ax}
  K.~Fukushima, F.~Gelis and L.~McLerran,
  Nucl.\ Phys.\  A {\bf 786}, 107 (2007)
%
\end{thebibliography}
\end{document}